\documentstyle[twocolumn,psfig,aps]{revtex}
\newcommand{\bramket}[3]{\left\langle\,{#1}\,\left|\,{#2}\,
            \right|\,{#3}\,\right\rangle}
\newcommand{\bqn}{\begin{eqnarray}}
\newcommand{\eqn}{\end{eqnarray}}
\newcommand{\beq}{\begin{equation}}
\newcommand{\eeq}{\end{equation}}
\newcommand{\CL}{{\cal L}}
\newcommand{\mbf}[1]{\mbox{\boldmath$#1$}}
\newcommand{\ovl}[1]{\overline{#1}}

\newcommand{\rw}{\rightarrow}

\begin{document}

\title{\bf Determination of $|V_{cb}|$ from exclusive decays in a
  relativistic quark model }

\author{M. Beyer}
\address{FB Physik Universit\"at Rostock, Universit\"atsplatz 1, 
18051 Rostock, FRG,\\
\sf FAX 0381-4982857, Email:beyer@darss.mpg.uni-rostock.de}

\maketitle

\begin{abstract}
  {\bf Abstract:} In the framework of a relativistic covariant
  Bethe-Salpeter model for the quark-antiquark system we present a
  renewed determination of the Cabbibo-Kobayashi-Maskawa matrix
  element $|V_{cb}|$. Complementing an earlier analysis applied to the
  whole decay spectrum for $B\rightarrow D^* e\nu$ we now also employ
  the ``zero-recoil method'' that uses the end point of the decay
  spectrum ($\omega=1$) and is suited for heavy-to-heavy transitions.
  The averaged experimental value extracted from the data at zero
  recoil, $|V_{cb}|{\cal F}(\omega=1)=0.0343\pm0.0015$, then leads to
  $|V_{cb}|=0.0360\pm 0.0016$. This value is somewhat larger than the
  one that uses the whole decay spectrum for the model analysis.  We
  also contrast this result to a nonrelativistic model and to recent
  experiments on the $B\rightarrow D e\nu$ semileptonic decay.
\end{abstract}
13.30.Ce, 
12.39.Ki, 
12.15.Hh, 
13.20.He 
\section{Introduction}

Within the standard model the extraction of the
Cabbibo-Kobayashi-Maskawa (CKM) matrix elements $|V_{cb}|$ (and
$|V_{ub}|$) is an outstanding topic of $B$-meson physics. Several ways
have been utilized that are summarized, e.g., by the Particle Data
Group~\cite{pdg96}. Presently, the value of $|V_{cb}|$ extracted from
inclusive decays is somewhat larger than from exclusive decays, e.g.,
in $B\rightarrow D^*e\nu$. 

A fruitful method to extract $|V_{cb}|$ from exclusive decays is to
reparameterize the decay data in such a way that they may be fitted by
a smooth (e.g., linear) function $|V_{cb}|{\cal F}(\omega)$ of
$\omega$, where $\omega=(m^2_B+m^2_{D^*}-q^2)/(2m_Bm_{D^*})$, and
$q^2$ is the 4-momentum transfer. Doing so it is possible to
extrapolate to the point of zero recoil of the $D^*$-meson, i.e.,
$\omega=1$, that is not directly measurable. This procedure is
particularly favored in the context of heavy quark expansion
(HQET)~\cite{isg90} but also useful to compare to other approaches
since in this context the notion of the whole decay spectrum is not
needed to extract $|V_{cb}|$. In HQET the value of ${\cal F}$ at zero
recoil is normalized up to corrections of order
$(\Lambda_{QCD}/m_{c,b})^2$ (where $m_{c,b}$ denotes the mass of the
$c$- or $b$-quark). However the required fitting and extrapolation
procedure leads to some errors, where the statistical error is under
control and presently in the order of $5\%$~\cite{san93}.

Alternatively, quark models have been proven very useful as they
provide not only predictions for ${\cal F}(\omega)$ for all $\omega$
and $|V_{cb}|$, but also numerous testable results for quite different
processes~\cite{wir85,isg88,jau90,fau92,sco95,zol95,gra96,mel96,che97}.
A general overview on bound state models for heavy hadron decay form
factors has been given by Ref.~\cite{ley97}. Other approaches to the
physics of heavy quark has profited from are QCD sum
rules~\cite{bra97} and lattice QCD~\cite{sac97}.

Since ${\cal F}(\omega)$ is known for a quark model one may ask for
the implications on the empirical value for $|V_{cb}|$, if the zero
recoil result is contrasted to the one obtained by using the whole
decay spectrum. Both methods are frequently used but not yet compared
to each other directly. In addition, relativistic quark models also
allow us to describe heavy-to-light transitions, in particular
$B\rightarrow \pi(\rho)$ important to determine $|V_{ub}|$, see, e.g.,
Ref.~\cite{stech}. In this sense the heavy-to-heavy transitions
provide an important test case and bear an impression of possible
model uncertainties.

In this context the merit of semileptonic $B\rightarrow D^{(*)}e\nu$
transitions may be considered twofold: As already mentioned they
provide an very good source to extract the
Cabbibo-Kobayashi-Maskawa (CKM) matrix element
$|V_{cb}|$ that has to be contrasted to inclusive and
nonleptonic decays.  On the other hand, weak decays (in general) 
provide important complementary information for QCD-motivated
modeling of the underlying quark structure of mesons (in general
hadrons). In addition, they may be considered useful to discuss the
different relativistic approximations used in this context.

We choose an approach utilizing the instantaneous Bethe-Salpeter
equation  to treat the $q\bar q$-system within a relativistically
covariant formalism~\cite{jres94}. The model is able to describe the
meson mass spectrum for low radial excitations. It has been applied to
the calculation of leptonic decays, viz.  decay constants,
$\gamma\gamma$-decays~\cite{mue94}, and to elastic form factors of
mesons~\cite{MR94} as well as to charmonium and
bottomonium~\cite{RM94}. Relativistic quark models have been
investigated, e.g., in Refs.~\cite{lag92,her93,tjo90,mur83,kal94}.

\section{The Bethe-Salpeter approach}

\subsection{Solving the Bethe-Salpeter equation}
The Bethe-Salpeter approach provides a consistent treatment of
two-body bound states as well as the coupling of an external field via
the Mandelstam formalism~\cite{sal51,man55}. In order to actually
solve the bound state problem several reasonable approximations are
necessary or practical: i) The quark propagators are assumed to be
free propagators irrespective of confinement that is introduced via a
confining kernel, ii) quark masses are assumed to be constant (i.e.
constituent quark mass) which is reasonable for heavy quarks, since
current quark masses and constituent quark masses needed for
reproducing the mesonic mass spectrum are rather close to each other,
iii) we utilize ladder approximation for the interaction kernel, and
iv) using an instantaneous interaction in addition leads to
computational advantages, as it provides RPA-type
equations~\cite{bil84} that can be solved by introducing an effective
Hamiltonian~\cite{lag92} in a formally covariant way. The specific
model used here, has been solved for the $q\bar q$-system in
\cite{jres94,mue94} and applied to a wide range of
phenomena~\cite{MR94,kle95} including the heavy quark
sector~\cite{zol95,RM94}. Details of the model may therefore be found
in the references given in the introduction. Here, I give a short
survey and summarize some results.

Within the approximations given above the $p_0$-integration in the
Bethe-Salpeter (BS) equation may be performed. The resulting Salpeter
amplitude in the rest frame of the bound state with mass \(M\) is the
given by
\begin{equation}
\Phi({\bf p})=\int \!\!\frac{dp^0}{(2\pi)}\,
\chi_P(p^0,{\bf p})|_{P=(M,{\bf 0})},
\label{eqn:amplitude}
\end{equation}
where $\chi_P(p^0,{\bf p})$ is the full Bethe-Salpeter
amplitude. 
Note, that the relative momentum $p=(p_0,{\bf p})$
appearing in Eq.~(\ref{eqn:amplitude}) may be written in a covariant
fashion~\cite{jres94,mue94}.

The resulting Salpeter equation is then given by
\begin{eqnarray}
\Phi({\bf p}) &=& 
\int \!\!\frac{d^3p'}{(2\pi)^3}\,
\frac{\Lambda^-_1({\bf p})\,\gamma^0\,
[V({\bf p},{\bf p}\,')\,\Phi({\bf p}\,')]
\,\gamma^0\,\Lambda^+_2(-{\bf p})}
{M+\omega_1+\omega_2} 
 \nonumber \\
&& -
\int \!\!\frac{d^3p'}{(2\pi)^3}\,
\frac{\Lambda^+_1({\bf p})\,\gamma^0\,
[V({\bf p},{\bf p}\,')\,\Phi({\bf p}\,')]
\,\gamma^0\,\Lambda^-_2(-{\bf p})}
{M-\omega_1-\omega_2}.
 \label{salpeter}
\end{eqnarray}
Here \(\omega_i=\sqrt{{\bf p}\,^2+m_i^2}\), and we introduce energy projection
operators \( \Lambda^{\pm}_i({\bf p}) = (\omega_i \pm
H_i({\bf p}))/(2\omega_i) \) in obvious notation,  where
\(H_i({\bf p})=\gamma^0(\mbf{\gamma}\cdot {\bf p}+m_i)\) is the standard
Dirac Hamiltonian (for details see, e.g., Refs.\cite{jres94,mue94}).

The dynamical input of the model is defined by a confinement plus one
gluon exchange (OGE) kernel, $V=V_C+V_G$.  Confinement is introduced
as a mixture of a scalar and a vector type kernel in the following way,
\begin{equation}
\left[V_C({\bf p},{\bf p}\,')\,\Phi({\bf p}\,')\right]
= {\cal V}^S_C(({\bf p}-{\bf p}\,')^2)\,[\Phi({\bf p}\,')
 - \gamma^0\,\Phi({\bf p}\,')\,\gamma^0].
\label{cc}
\end{equation}
 Due to the instantaneous
approximation it is possible to introduce the same spatial dependence
(in the rest system of the meson) as used in the nonrelativistic case,
viz. in co-ordinate space,
\begin{equation}
{\cal V}_C^F(r) = a_c+b_c r.
\label{vconf}
\end{equation}
The mixture of a scalar and a vector spin structure has been
introduced in order to give an improved description of the spin orbit
splitting. Other mixtures have been advocated in the literature, and
also anomalous tensor-type confinement has been discussed, see, e.g.,
Ref.~\cite{fau92}. However, the consequences concerning, e.g., the
mass spectrum or Regge behavior have not been studied yet.

For the OGE kernel, we chose the Coloumb gauge for the gluon
propagator. This way it is possible to retain a covariant formulation
within an instantaneous treatment of the Bethe-Salpeter equation, and it allows
to substitute \(q^2\) by \(-{\bf q}^{\,2}\). 
 The OGE kernel then reads \cite{tjo90,mur83}
\begin{eqnarray}
\lefteqn{\left[V_G({\bf p},{\bf p}\,')\,\Phi({\bf p}\,')\right]
= {\cal V}_G(({\bf p}-{\bf p}\,')^2)} \nonumber \\ 
&& \times \left[
\gamma^0 \Phi({\bf p}\,') \,\gamma^0
-\frac{1}{2}\,\left(
\mbf{\gamma} \Phi({\bf p}\,')\, \mbf{\gamma} +
(\mbf{\gamma}\hat{x}) \Phi({\bf p}\,')\, (\mbf{\gamma}\hat{x})\,
\right)\,\right],
\label{Couleq}
\end{eqnarray}
with the operator $\hat x={\bf x}/|{\bf x}|$, and
\begin{equation}
{\cal V}_G({\bf q}^{\,2}) = 
\pi\,\frac{4}{3}\,\frac{\alpha_s({\bf q}^{\,2})}{{\bf q}^{\,2}},
\label{vgluon}
\end{equation}
where $\alpha_s(q^2)$ is introduced as ``running'' coupling as
discussed in Refs.~\cite{zol95,jres94,mue94}.

To solve the Salpeter equation numerically, Eq. (\ref{salpeter}) is
rewritten as an eigenvalue problem (RPA-equations), see, e.g.,
Ref.~\cite{jres94,lag92}.  This way it is possible to utilize the
variational principle to find the respective bound states. To this end
the Salpeter amplitude $\Phi$ is expanded into a reasonable large
number of basis states used as a test function. As a suitable choice
of basis states we have taken Laguerre polynomials and found that
about ten basis states lead to sufficient accuracy, see
also Refs.~\cite{jres94,mue94}.

For completeness, the parameters of the model given in Ref.~\cite{zol95}
are shown in Table~\ref{parameters}.  These are the quark masses,
the offset \(a_c\) and slope \(b_c\) of the confinement interaction
Eq.~(\ref{vconf}) and the saturation value
\(\alpha_{sat}=\alpha_{s}({\bf q}^2\rightarrow 0)\).  They are
determined to give a good overall description of the meson mass
spectrum (heavy and light mesons as well as charmonium and
bottomonium)~\cite{zol95,jres94,mue94,MR94,RM94,kle95}.

\subsection{Current matrix elements}

Semileptonic decays are treated in current-current approximation.  For
a transition $b\rw c$ the Lagrangian is given by
\begin{equation}
\label{lagrangian}
\CL_{cb}=\frac{G_F}{\sqrt{2}}\;V_{cb}\;\;h_{cb}^\mu\;j_\mu,
\end{equation}
with the CKM matrix element $V_{cb}$ and the
Fermi constant $G_F$.
The leptonic  $j_\mu$ and hadronic currents $h_{cb}^\mu$ are
defined by 
\begin{eqnarray}
j_\mu&= &{\ovl \ell} \gamma_\mu (1-\gamma_5) \nu_\ell,\\
h_{cb}^\mu&=&{\ovl c} \gamma^\mu (1-\gamma_5) b.
\label{curhad}
\end{eqnarray} 
The relevant transition amplitudes 
\begin{equation}
\bramket{D^{(*)}} {\bar c \gamma^\mu (1-\gamma_5) b}{B}
\end{equation}
for $B\rw D$ and $B\rw D^*$ of the hadronic current can be decomposed
due to the Lorentz structure of the current, thus introducing form
factors.  A standard representation of the form factors is given in
terms of $F_0(q^2)$, $F_1(q^2)$ (or $f_+(q^2)$, $f_-(q^2)$) for
$0^-\rw 0^-$ transitions, and $V(q^2)$, $A_0(q^2)$, $A_1(q^2)$,
$A_2(q^2)$ for $0^- \rw 1^-$ transitions~\cite{koe88}.  The exact
definitions, and further references have been given, e.g., in
Ref.~\cite{res94}. Note that $m_\ell^2\leq q^2\leq
q_{max}^2=(m_B-m_{D^{(*)}})^2$ due to kinematical reasons. Helicity
amplitudes $H_\pm$ and $H_0$ in terms of the above form factors have
been given by K\"orner and Schuler in a series of papers~\cite{koe88}
and are compiled by the particle data group~\cite{pdg96}. Using the
helicity amplitudes the formulas for the decay spectrum used here
is given in Ref.~\cite{koe88} and will no be repeated here. The
respective decay rates into specific helicity states $\Gamma_\pm$,
$\Gamma_0$ are also given in the literature, see, e.g.,
Ref.~\cite{pdg96}.

\begin{figure}[pthb]
\centering
  \psfig{figure=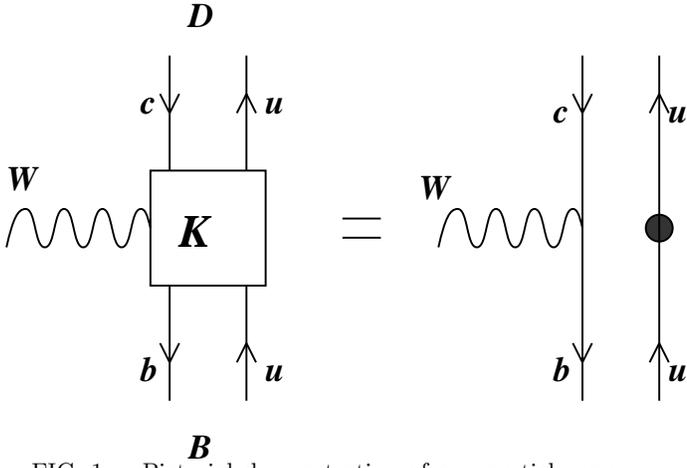,width=0.5\textwidth}
\caption{\label{kernel}
Pictorial demonstration of one particle approximation of the
irreducible interaction kernel, filled
circle denotes the inverse quark propagator } 
\end{figure}

To determine the form factors from the model, we follow the general
prescription by Mandelstam~\cite{man55}, see, e.g., Ref.~\cite{lur68}
for a textbook treatment.  The lowest order contribution (relativistic
impulse approximation) to the current (sometimes referred to as
triangle graph) is given in Fig.~\ref{kernel}, and written as
(consider, e.g., the anti-quark current, flavor indices suppressed)
\begin{eqnarray}
\label{transitioncurrent}
\lefteqn{\left\langle\,D^*, P_{D^*}\,\left|\, 
h^\mu_{cb}(0)\, 
\right|\,B, P_B\,\right\rangle = -
 \;\int\!\!\frac{d^4p}{(2\pi)^4}}\\
&&\qquad{\rm tr} \; \left\{
 \bar{\Gamma}_{P_{D^*}}(p-q/2)\;{S^F_{{\ovl q}'}}(P_B/2+p-q)\;\right.
\nonumber\\
&&\qquad\times \gamma^{\mu}(1-\gamma_5)\;\nonumber\\
 &&\qquad\left. \times{S^F_{\ovl q}}(P_B/2+p)\;
 \Gamma_{P_B}(p)\;{S^F_q}(-P_B/2+p)\right\}
 \nonumber
\end{eqnarray} 
where \(p\) and \(p'\) denote the relative momenta of the incoming and
outgoing \(q\bar{q}\) pair, \(q=P_{D^*}-P_B\) is the momentum transfer. The
quark Feynman propagator is denoted by $S^F_q$. The Dirac coupling to
point-like particles is consistent with the use of free quark
propagators. In Eq.~(\ref{transitioncurrent}) the amputated
Bethe-Salpeter amplitude or vertex function $\Gamma_P(p)$ is given by
\begin{equation}
 \Gamma_P(p)  := 
[S^F_{q}(p_{q})]^{-1} \,\chi_P(p)\;[S^F_{\ovl q}(-p_{\ovl q})]^{-1}.
\end{equation}
It may be computed in the rest frame from the equal time amplitude
$\Phi({\bf p})$ using the  Bethe-Salpeter equation
\begin{eqnarray} 
  \Gamma_P(p)\left|_{_{P=(M,{\bf 0}\,)}}\right.  \; \equiv\;
\Gamma({\bf p}\,) \; =\;
  -i\! \int\!\! \frac{d^3p'}{(2\pi)^4}
  \left[ V({\bf p},{\bf p}\,')\Phi({\bf p}\,')\right].
\label{vert}
\end{eqnarray} 
Finally, using the Lorentz transformation properties of the field
operators that define the Bethe-Salpeter amplitude~\cite{lur68},
 we can calculate the Bethe-Salpeter
amplitude in any reference frame via
\begin{equation}
  \chi_P(p) = \;
  S_{\Lambda_P}^{}\;\;\chi_{(M,{\bf 0})}(\Lambda_P^{-1}p)\;\;
  S_{\Lambda_P}^{-1}.
\label{boo}
\end{equation}
where   $\Lambda_P$ is the pure Lorentz boost, and $S_{\Lambda_P}$ the
corresponding transformation matrix for Dirac spinors. 

Due to the reconstruction of the full Bethe-Salpeter amplitude sketched
above, the transition matrix element Eq.~(\ref{transitioncurrent}) is
manifestly covariant.

\subsection{Form factors}

The analysis of experimental data on heavy-to-heavy transitions now
widely uses the notion of heavy quark expansion~\cite{pdg96}.
Following Ref.~\cite{neu94} we for one introduce the ratios
$R_1(\omega)$ and $R_2(\omega)$,
\begin{eqnarray}
R_1(\omega)&\equiv&\left[1-\frac{q^2}{(m_B+m_{D^*})^2}\right]
\frac{V(q^2)}{A_1(q^2)},\\
R_2(\omega)&\equiv&\left[1-\frac{q^2}{(m_B+m_{D^*})^2}\right]
\frac{A_2(q^2)}{A_1(q^2)},
\end{eqnarray}
where $\omega=(m_B^2+m_{D^*}^2-q^2)/(2m_Bm_{D^*})$. 

For $B\rightarrow
D^*e\nu$ decays the standard form factors may then be related to the
 ones used for the heavy quark expansion by~\cite{neu94}
\begin{eqnarray}
A_1(q^2)&=&\kappa_{BD^*}\,
\left[1-\frac{q^2}{(m_B+m_{D^*})^2}\right]
h_{A_1}(\omega),\\
A_2(q^2)&=&\kappa_{BD^*}\,R_2(\omega)h_{A_1}(\omega),\\
V(q^2)&=&\kappa_{BD^*}\,R_1(\omega)h_{A_1}(\omega).
\end{eqnarray}
where $\kappa_{BD^*}=(m_B+m_{D^*})/(2\sqrt{m_Bm_{D^*}})$.
In the heavy quark mass limit ($m_{c,b}\rightarrow \infty$)
\begin{eqnarray}
h_{A_1}(\omega)&\rightarrow&\xi(\omega),\label{ffisgw}\\
R_{1,2}(\omega)&\rightarrow&1.
\end{eqnarray}
where $\xi(\omega)$ is a universal function known as Isgur-Wise
function~\cite{isg90}.  

For the case $B\rightarrow De\nu$ the standard form factors are
related to the heavy quark form factors via~\cite{neu94},
\begin{eqnarray}
f_+(q^2) &=&\kappa_{BD}^+h_+(\omega)-\kappa_{BD}^-h_-(\omega),\\
f_-(q^2) &=&\kappa_{BD}^+h_-(\omega)-\kappa_{BD}^-h_+(\omega),
\end{eqnarray}
where $\kappa_{BD}^\pm=(m_B\pm m_D)/(2\sqrt{m_Bm_D})$. In the heavy
quark mass limit $h_+(\omega)\rightarrow\xi(\omega)$ and
$h_-(\omega)\rightarrow 0$.

\begin{figure}[pthb]
 \leavevmode
 \centering
  \psfig{figure=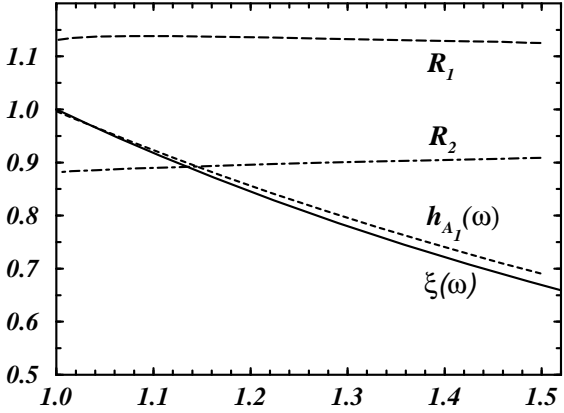,width=0.5\textwidth,angle=270}
\caption{\label{fig:isgurwise}
  Form factors $R_1$ (long-dashed), $R_2$ (dashed-dotted), and
  $h_{A_1}$ for the $B\rightarrow D^*$ transition. The Isgur-Wise
  function $\xi$ of heavy quark mass limit is shown as solid line.
  Radiativ corrections not included here.}
\end{figure}

In the model approach used here the heavy quark mass limit has been
performed numerically by multiplying $m_{c,b}$ with a large factor and
keeping all other parameters as given in Table~\ref{parameters}. To
evaluate the transition matrix elements Eq.~(\ref{transitioncurrent})
the meson amplitudes are then calculated by diagonalizing the
eigenvalue problem with the large quark masses.  Due to numerical
reasons the heavy quark masses cannot be chosen too large. The
function resulting from this numerical limiting procedure is then
defined to be the Isgur Wise function $\tilde\xi(\omega)$ of the
Bethe-Salpeter-model, where $\tilde\xi(1)=1.00$ within 0.1\%. This
function is shown as a solid line in Fig.~\ref{fig:isgurwise}. Note
that at this stage $\tilde\xi$ does not include radiative corrections
that will be given below. In the same fashion the ratios $R_1$ and
$R_2$ tend to unity within less than 0.1\% when numerically increasing
the heavy quark masses.

For finite masses the experimental ratios $R_{1,2}$ assuming constant
values have recently been extracted by CLEO~\cite{san93}. The
latest values are~\cite{gol97},
\begin{eqnarray}
R_1 &=& 1.24\pm0.26\pm0.12,\\
R_2&=& 0.72\pm 0.18\pm 0.07,
\end{eqnarray}
that have to be contrasted to the long-dashed ($R_1$) and
dashed-dotted ($R_2$) curve shown in Fig.~\ref{fig:isgurwise}. The
model ratios vary slowly by roughly 10\% over the whole
$\omega$ range, which is smaller than the experimental error.

\section{Results}
Utilizing the Bethe-Salpeter model to describe mesons as $q\bar q$
states the exclusive decay spectra for $B\rightarrow D^*e\nu$ and
$B\rightarrow De\nu$ have been calculated and compared to the
experimental data. Earlier the CKM matrix element $|V_{cb}|$ has been
determined by a least squared fit to the whole spectrum of
$B\rightarrow D^*e\nu$~\cite{zol95}. We have now redone this analysis
on the basis of the improved data and also included the recently
measured $B\rightarrow De\nu$ decay spectrum. In addition, we present
a new analysis for this model approach at zero recoil of $B\rightarrow
D^*e\nu$ that is commonly used for heavy-to-heavy transitions to
extract $|V_{cb}|$. This enables us to compare the difference between
the energy dependent and the zero-recoil analysis quantitatively, at
least for the Bethe-Salpeter model discussed here and the
nonrelativistic approach given earlier~\cite{res94}.

\subsection{$B\rightarrow D^*e\nu$ decay}

We now turn to the extraction the CKM matrix element $|V_{cb}|$. The
differential decay rate for $B\rightarrow D^*e\nu$ is given
by~\cite{pdg96,neu91},
\begin{eqnarray}
\frac{d\Gamma_{D^*}}{d\omega} &= 
&\frac{G_F^2}{48\pi^3}m_{D^*}^3(m_B-m_{D^*})^2\nonumber\\
&&\times\sqrt{\omega-1}(\omega+1)^3
|V_{cb}|^2{\cal F}_{D^*}^2(\omega)\nonumber\\
&&\times\left[1+\frac{4\omega}{\omega+1}
\frac{m_B^2+m_{D^*}^2-2\omega m_B m_{D^*}}
{(m_B-m_{D^*})^2}\right].
\end{eqnarray}
The formula is written in a way that ${\cal F}_{D^*}(\omega)$ reduces
to ${\cal
  F}_{D^*}(\omega)\rightarrow\eta_A\tilde\xi(\omega)=\xi(\omega)$ in
the heavy quark mass limit, where $\eta_A$ denotes the radiative
corrections~\cite{neu94,cza96}.  For finite masses the function ${\cal
  F}_{D^*}(\omega)$ contains all the symmetry breaking effects.

\begin{figure}[pthb]
 \leavevmode
\centering
  \psfig{figure=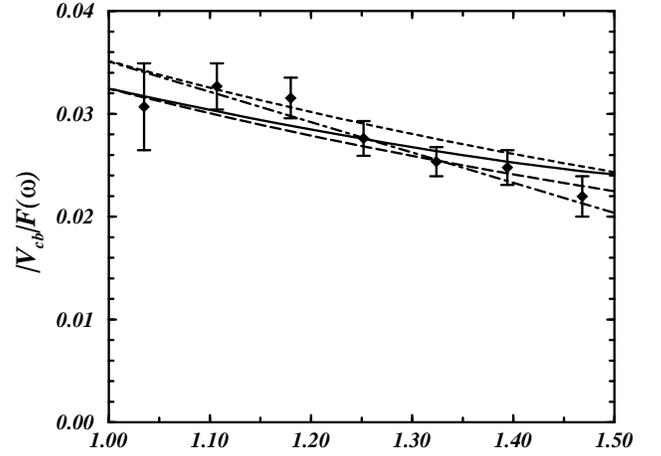,width=0.5\textwidth,angle=270}
\caption{\label{fig:cleo}
  $|V_{cb}|{\cal F}(\omega)$ as a function of $\omega$ for
  $B\rightarrow D^* e\nu$ decays. Data points from CLEO and their
  linear fit (with $c=0$, dashed-dotted line). Model best fit is
  displayed as solid line. Long-dashed with the same value for
  $|V_{cb}|$ at $\omega=1$ but $R_{1,2}$ constant. Dashed line uses
$V_{cb}$ from CLEO analysis and $R_{1,2}$ constant (i.e. zero-recoil
method) for comparison.}
\end{figure}

Experiments are given in a way that all well known factors are divided
out in the decay rate and only $|V_{cb}|{\cal F}_{D^*}(\omega)$ is
left over. The corresponding data points of a recent CLEO measurement
are shown in Fig.~\ref{fig:cleo}.  The result is particularly smooth
and may be fitted by a linear curve. The fit to the data done by the
CLEO collaboration is also shown in Fig.~\ref{fig:cleo} as
dashed-dotted line.  Other lines reflect the model results utilizing
different assumptions. The solid line is calculated using the exact
formula for the decay rate as given, e.g., in~\cite{pdg96,koe88}
(i.e., with $R_{1,2}$ $\omega$-dependent) divided by the same factor
as the experiments are. The CKM matrix element $|V_{cb}|$ is then
determined by a least squared fit to all data points. Radiative
corrections have been included in the dominant form factor
$h_{A_1}(\omega)$, expressed as an overall factor
$\eta_A$~\cite{neu94}. Unlike earlier estimates that imply a
correction of approx. 1\%~\cite{neu94} (i.e., smaller than the model
uncertainty and therefore neglected in the earlier
analysis~\cite{zol95}) a recent two loop calculation leads to
substantial value of $\eta_A=0.960\pm 0.007$\cite{cza96}, which has to
be included into the analysis. The result is $|V_{cb}|=0.0339\pm
0.0010$.  To see the model dependence of the different analyses used
in this context we now take ${\cal
  F}_{D^*}(\omega)=\eta_A\,h_{A_1}(\omega)$, $R_1$ and $R_2$ constant,
i.e., $R_1(1)$, $R_2(1)$, and this CKM matrix element that leads to
the long-dashed line shown in Fig.~\ref{fig:cleo}. It is obvious that
the curve slightly deviates from the solid one that includes the
$\omega$-dependence of $R_1$ and $R_2$. For the same form factor
however using the value of $|V_{cb}|{\cal F}_{D^*}(\omega=1)$ from the
CLEO fit is given by the short-dashed curve. The resulting CKM matrix
element is obviously larger by $\simeq 8\%$.

The function ${\cal F}_{D^*}(\omega)$ that leads to Fig.~\ref{fig:cleo}
can be approximated by a quadratic fit to
\begin{equation}
{\cal F}_{D^*}^{(2)}(\omega)={\cal F}_{D^*}(1)\left(1-\rho_{A_1}^2(\omega-1)
+c(\omega-1)^2\right),
\end{equation}
with the parameters $\rho_{A_1}^2$ and $c$. The slope of ${\cal
  F}_{D^*}(\omega)$ extracted by CLEO~\cite{san93} assuming a
linear dependence on $\omega$ ($c=0$) is $ \rho_{A_1}^2 = 0.84\pm 0.13
\pm 0.08$, and show as dashed-dotted line in Fig.~\ref{fig:cleo}. The
respective parameters for the quadratic fit of the curve discussed are
shown in Tab.~\ref{tab:slope}.

Within the notion of the heavy quark effective theory the form factor
${\cal F}_{D^*}(\omega=1)$ can be expanded into orders of
$\Lambda_{QCD}/m_{c,b}$. Since symmetry breaking effects in
semileptonic decays are of second order only~\cite{luk90},
lowest order terms are usually written as
\begin{equation}
{\cal  F}_{D^*}(\omega=1)=\eta_A(1+\delta_{1/m^2}),
\end{equation}
where $\delta_{1/m^2}$ has to be determined.  Corrections vary from
$-\delta_{1/m^2}=(3\pm 2)\%$~\cite{fal93,man94} to
$-\delta_{1/m^2}=(5.5\pm 2.5)\%$~\cite{shi95}. A recent discussion
and appreciation of the different approaches is given by
Martinelli~\cite{mat96}.  The relativistic model discussed here
leads to value of $-\delta_{1/m^2}=0.5\%$.

The values for ${\cal F}_{D^*}(\omega=1)|V_{cb}|$ that have been extracted by
different experiments are quite
consistent and lead to an overall fit of~\cite{mat96,gol97} 
\begin{equation}
{\cal F}_{D^*}(\omega=1)|V_{cb}|=0.0343\pm 0.0015
\end{equation}
for the empirical slope parameter given in the last line of
Tab.~\ref{tab:slope}.  Using this value and the radiative corrections
given above we extract the CKM matrix element $|V_{cb}|$ for the
relativistic Bethe Salpeter model to be
\begin{eqnarray}
|V_{cb}|&=&0.0360\pm  0.0016\quad\mbox{zero-recoil,}\\
|V_{cb}|&=&0.0339\pm  0.0010\quad\mbox{full spectrum}.
\end{eqnarray}
This is the main result that shows the potential model dependence of the
zero-recoil method that adds to the statistical uncertainty.
A similar renewed analysis for the nonrelativistic model given
before~\cite{res94} now leads to $|V_{cb}|=0.037\pm 0.002$ that is
larger by 7\% compared to the spectrum dependent analysis. 

\subsection{$B\rightarrow De\nu$ decay}

\begin{figure}[pthb]
 \leavevmode
\centering
  \psfig{figure=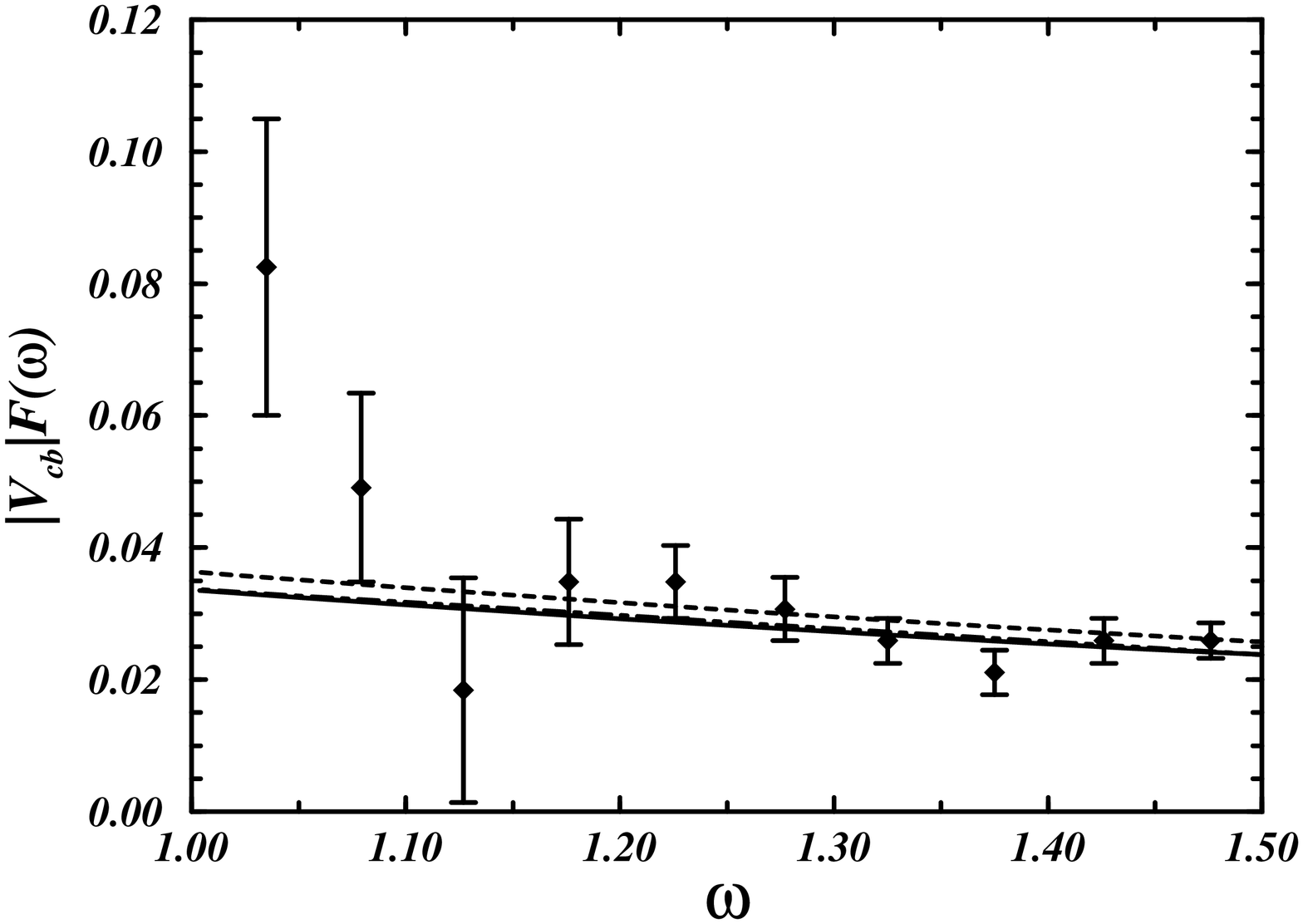,width=0.5\textwidth,angle=270}
\caption{\label{fig:cleoBD}
  $|V_{cb}|{\cal F}(\omega)$ as a function of $\omega$ for
  $B\rightarrow D e\nu$ decays. Data points from CLEO and their linear
  fit (with $c=0$, dashed-dotted line). Model best fit is displayed as
  solid line, same $V_{cb}$ as in Fig.~\ref{fig:cleo}.  }
\end{figure}

The differential decay rate for the $B\rightarrow De\nu$ is given
by~\cite{neu91},
\begin{eqnarray}
\frac{d\Gamma}{d\omega} &= 
&\frac{G_F^2}{48\pi^3}m_{D}^3(m_B+m_{D^*})^2\nonumber\\
&&\times(\omega-1)^{3/2}|V_{cb}|^2{\cal F}_D^2(\omega)
\end{eqnarray}
where again ${\cal F}_D(\omega)$ reduces to the Isgur Wise function
in the heavy quark mass limit. Recent experimental data for the
relevant part of the spectrum $|V_{cb}|{\cal F}_D(\omega)$ are shown
in Fig.~\ref{fig:cleoBD}. The linear fit to the data given by the CLEO
collaboration is also shown (as a dashed-dotted line). The model
results utilizing the full spectrum dependent analysis leads to the
solid line in Fig.~\ref{fig:cleoBD}. For comparison the result of the
zero recoil method utilized in the previous paragraph is also shown.
The radiative corrections have been assumed to be in the same order as
in the $B\rightarrow D^*e\nu$ transition. Obviously the model is
capable to provide a good description of the experimental data.

\section{Summary and Conclusion}

We have analyzed the exclusive decay rates of $B\rightarrow D^*e\nu$
and $B\rightarrow De\nu$ within a relativistic constituent quark
model.  The interaction kernel has been taken instantaneous. This way
the Bethe Salpeter equation reduces to a Salpeter equation as given in
Eq.~(\ref{salpeter}). The interaction consists of a one gluon exchange
evaluated in the Coulomb gauge and a linear confinement given in
co-ordinate space. The model parameters have been fixed to describe
the mass spectrum of all observed mesons (not only heavy mesons) in a
satisfactory manner~\cite{jres94,mue94,MR94,RM94}. The interaction
current to describe the weak decay process has been introduced via the
Mandelstam formalism. To this end the (instantaneous) amplitude has
been reconstructed using the Lorentz transformation properties of the
field operators.

The only parameter left to describe the exclusive decay spectra is the
CKM matrix element $|V_{cb}|$. For $B\rightarrow D^*e\nu$ two methods
have been compared. One uses the complete spectrum, viz. the
functional dependence of ${\cal F}_{D^*}$, emerging from the quark
model. The CKM matrix element is then fixed by a least squared fit.
The other analysis utilizes the ``zero recoil'' method used in the
context of heavy quark expansion. Here only the empirical value of
$|V_{cb}|{\cal F}_{D^*}(1)$ is used that is gained from an
extrapolation of the experimental data (e.g. by a linear fit) to the
zero recoil point $\omega=1$ and $R_1$, $R_2$ assumed constant. The
CKM matrix element $|V_{cb}|$ is then extracted using a singular value
of the model at ${\cal F}_{D^*}(1)$.  Clearly, this method is not
consistent with the underlying model, but widely used to extract
$|V_{cb}|$ from the zero recoil point. Comparing the full model result
to the zero-recoil result leads to different values for the CKM matrix
elements $|V_{cb}|$ by approximately $8\%$. In view of this result it
seems obvious that this kind of model uncertainty steaming from the
different treatment of the $\omega$-dependence of the spectrum may
show up in the determination of $|V_{cb}|$ besides the statistical
error. This result may also be relevant for other more ``model
independent'' analyses.

For comparison the  $B\rightarrow De\nu$ decay has been calculated for
the different assumptions and also leads to a good overall description of
the experimental spectrum. 

\section{Acknowledgment}
I would like to thank G. Z\"oller for his previous contributions to
this work.  I am grateful to D. Melikhov for valuable comments on the
manuscript and to T. Mannel for his interest.  Also I would like to
thank R. Faustov for discussions on some general issues.

\begin{table}[phtb]
\caption{Parameters of the BS model. }
\label{parameters} 
\[
\begin{array}{cccccccc}
\hline\hline
m_{u,d} &m_s   &m_c     &m_b     &a_c     &b_c       &r_0     
&\alpha_{sat}\\
\mbox{[GeV]}  &\mbox{[GeV]}  &\mbox{[GeV]}  &\mbox{[GeV]}
&\mbox{[GeV]} &\mbox{[GeV/fm]}& \mbox{[fm]} &\\[1ex]
\hline
\vspace{1ex}
0.200 &0.440 &1.738 &5.110 &-1.027 &1.700 &0.1  &0.391\\
\hline\hline
\end{array}
\]
\end{table}
\begin{table}[pthb]
\caption{\label{tab:slope} Parameters for a quadratic fit to the CLEO
  data and the BS model. First line in heavy quark mass limit, other
  lines for physical masses.}
\begin{center}
\begin{tabular}{ccl}
$\rho^2_{A_1}$&c&model\\
\hline
0.83   &0.34 & for $\xi_{BS}(\omega)$\\
0.76   &0.30 & for $h_{A_1} (\omega)$\\  
0.69   &0.34 & for ${\cal F}  (\omega)$\\
$0.92\pm0.64\pm0.40$&$0.15\pm1.24\pm0.90$& CLEO data
\end{tabular}
\end{center}
\end{table}



\begin{thebibliography}{99}
\bibitem{pdg96} R.M. Barnett {\it et al.}, Phys. Rev. {\bf D 54}, 1 (1996), 
     and 1997 off-year partial update for the 1998 edition available on 
     the PDG WWW pages (URL: http://pdg.lbl.gov/).
\bibitem{isg90}
N. Isgur, M. B. Wise, Phys. Lett. {\bf B 232}, (1989) 113; {\bf B
  237}, 527 (1990).
\bibitem{san93} CLEO Collaboration,
S. Sanghera {\it et al.}, Phys. Rev. {\bf D 47}, 791 (1993);
B. Barish {\it et al.},
  Phys. Rev. {\bf D 51}, 1014 (1995); J.E. Duboscq
  {\it et al.} Phys. Rev. Lett. {\bf 76} 3898 (1996).
\bibitem{wir85} M. Wirbel, B. Stech, M. Bauer, Z. Phys. {\bf C 29},
  637 (1985).
\bibitem{isg88} N. Isgur, D. Scora, B. Grinstein, M.B. Wise,
  Phys. Rev. {\bf D 39}, 799 (1989).
\bibitem{jau90} W.Jaus, Phys. Rev. {\bf D 41} 3394 (1990), ebd. {\bf D
    53}, 1349 (1996).
\bibitem{fau92} V.O. Galkin, A.Yu. Mishurov, R.N. Faustov, Sov. J. Nucl.
Phys. {\bf 55} (1992) 1207, R.N. Faustov, V.O. Galkin, A.Yu. Mishurov, J.
Nucl. Phys. {\bf 55} (1992) 1080 (english: JINR print Dubna, Russia
E2-91-451), R.N. Faustov, V.O. Galkin, A.Yu. Mishurov, 
Phys. Lett. {\bf B 356}, 516 (1995); {\it ibid.} {\bf B367} E391 (1996);
 R.N. Faustov, V.O. Galkin, Z. Phys. {\bf C 66}, 119 (1995).
\bibitem{sco95} D. Scora, Nathan Isgur, Phys. Rev. {\bf D 52}, 2783
  (1995). 
\bibitem{zol95} G. Z\"oller, S. Hainzl, C.R. M\"unz, M. Beyer,
  Z. Phys. {\bf C 68}, 103 (1995); G. Z\"oller, {\it diploma thesis} (in
  German, unpublished, Bonn 1994).
\bibitem{gra96} I.L. Grach, I.M. Narodetskii, S. Simula,
  Phys. Lett. {\bf B 385}, 317 (1996). 
\bibitem{mel96} D. Melikov,  Phys. Lett. {\bf B 394}, 385 (1997),
  {\it ibid.} {\bf B 380} (1996), Phys. Rev. {\bf D 53}, 2460 (1996). 
\bibitem{che97} H.-Y. Cheng, C.-Y. Cheung, C.-W. Hwang,
  Phys. Rev. {\bf D 55}, 1559 (1997). 
\bibitem{ley97} A. Le Yaouanc invited plenary
  talk on {\it IVth International Workshop on Progess in Heavy Quark
    Physics} held in Rostock 1997, to be published in the proceedings
  (eds. M. Beyer, T. Mannel, H. Schr\"oder, Rostock University Press,
  1998).  
\bibitem{bra97} V. Braun invited plenary talk on {\it IVth International
    Workshop on Progess in Heavy Quark Physics} held in Rostock 1997, to be
  published in the proceedings (eds. M. Beyer, T. Mannel,
  H. Schr\"oder, Rostock University Press, 1998).
\bibitem{sac97} C. Sachrajda invited plenary talk on {\it IVth International
    Workshop on Progess in Heavy Quark Physics} held in Rostock 1997, to be
  published in the proceedings (eds. M. Beyer, T. Mannel,
  H. Schr\"oder, Rostock University Press, 1998).
\bibitem{stech} B. Stech,   Z. Phys. {\bf C 75}, 245 (1997). 
\bibitem{jres94} J.\,Resag, C.R.\,M\"unz, B.C.\,Metsch, H.R.\,Petry,
  Nucl. Phys.  {\bf A 578} 397 (1994).
\bibitem{mue94} C.R.\,M\"unz, J.\,Resag, B.C.\,Metsch, H.R.\,Petry, 
Nucl. Phys. {\bf A 578}, 418 (1994).
\bibitem{MR94} C.R.\,M\"unz, J.\,Resag, B.C.\,Metsch, H.R.\,Petry,
  Phys. Rev. {\bf C52}, 2110 (1995).
\bibitem{RM94} J.\,Resag, C.R.\,M\"unz, Nucl. Phys. {\bf A 590}, 735
  (1995). 
\bibitem{lag92}
J.F. Laga\"e, Phys. Rev. {\bf D 45} 305, 317 (1992).
\bibitem{her93}
H. Hersbach, Phys. Rev. {\bf A 46} 3657 (1992), {\bf D 47} 3027 (1993).
\bibitem{tjo90} E.\,Hummel, J.\,Tjon, Phys. Rev. {\bf C 42}, 423 (1990);
G.\,Rupp, J.A.\,Tjon, Phys. Rev. {\bf C 41}, 472 (1990);
P.C.\,Tiemeijer, J.A.\,Tjon: Phys. Lett. {\bf B 277}, 38 (1992); 
Phys. Rev. {\bf C49} 494 (1993), {\bf C 48} 896 (1994).
\bibitem{mur83} T.\,Murota, Progr. Theor. Phys. {\bf 69}, 181
  (1983), ibid. 1498 (1983).
\bibitem{kal94}
Yu.L. Kalinovsky, C. Weiss: Z. Phys. {\bf C 63}, 275 (1994).
\bibitem{sal51} E.E. Salpeter, H.A. Bethe, Phys. Rev. {\bf 84}, 1232
  (1951). 
\bibitem{man55} S.\,Mandelstam, Proc. Roy. Soc. {\bf 233}, 248 (1955)
\bibitem{bil84} A. Bilal, P. Schuck, Phys. Rev. {\bf D 31}, 2045
  (1985).  
\bibitem{kle95} E. Klempt, B.C. Metsch, C.R. M\"unz, H.R. Petry,
Phys. Lett. {\bf B361}, 160 (1995), W.I. Giersche, C.R. M\"unz, 
Phys. Rev. {\bf C 53}, 2554 (1996), C.R. M\"unz, Nucl. Phys. {\bf A
  609}, 364 (1996).
\bibitem{koe88} J.G. K\"orner, G.A. Schuler, Z.
  Phys. {\bf C 38}, 511 (1988) 511 [E {\bf C 41}, 690 (1989)]; {\bf C
  46}, 93 (1990).
\bibitem{neu94} M. Neubert, Phys. Rep. {\bf
    245}, 259 (1994).  
\bibitem{gol97} CLEO collaboration, G. Gollin invited plenary talk
  on {\it IVth International Workshop on Progess in Heavy Quark Physics}
  held in Rostock 1997, to be published in the proceedings (eds. M.
  Beyer, T. Mannel, H. Schr\"oder, Rostock University Press, 1998).
\bibitem{res94} S. Resag, M. Beyer, Z.
  Phys. {\bf C 63} 121 (1994).
\bibitem{lur68} D.\,Lurie, {\it Particles and Fields},
  (Interscience Publishers, 1968). 
\bibitem{neu91} M. Neubert, Phys. Lett. {\bf B 264}, 455 (1991).
\bibitem{cza96} A. Czarnecki, Phys. Rev. Lett. {\bf 76} 4124 (1996).
\bibitem{luk90} M.E. Luke, Phys. Lett. {\bf B 252}, 447 (1990).
\bibitem{fal93} A.F. Falk, M. Neubert, Phys. Rev. {\bf D47} 2965, 2982
  (1993).  
\bibitem{man94} T. Mannel, Phys. Rev., {\bf D 50} 428
  (1994).  
\bibitem{shi95} M.A. Shifman, N.G. Uraltsev, A. Vainshtein,
  Phys. Rev. {\bf D51}, 2217 (1995).  
\bibitem{mat96} G. Martinelli,
  Nucl. Instrum. Meth. {\bf A 384}, 241 (1996).


\end{thebibliography}
\end{document}